\documentclass[aps,prx,a4paper,notitlepage,reprint]{revtex4-2}

\usepackage{amsmath,amssymb,amsfonts} 
\usepackage[cal=pxtx,scr=boondoxo, bb=pazo,scrscaled=1.05]{mathalfa}                                   

\usepackage{upgreek}
\usepackage{mathtools} 
\usepackage{color}

\usepackage{graphicx,float}
\usepackage[colorlinks,
linkcolor=red,
citecolor=blue,
urlcolor=red]{hyperref}

\usepackage{bm}
\usepackage{soul}
\usepackage{pgf}
\usepackage{siunitx} 
\usepackage{transparent}

\usepackage{xspace}

\renewcommand{\t}[1]{\mathrm{#1}}

\newcommand{\figref}[1]{Fig.~\ref{#1}}
\renewcommand{\eqref}[1]{Eq.~(\ref{#1})}
\newcommand{\secref}[1]{Sec.~\ref{#1}}

\newcommand{\ket}[1]{\vert{#1}\rangle}
\newcommand{\bra}[1]{\langle{#1}\vert}

\begin{document}
\title{Squeezed light from an oscillator measured at the rate of oscillation}

\author{Christian Bærentsen}

\author{Sergey A. Fedorov}
\email{sergey.fedorov@nbi.ku.dk}

\author{Christoffer Østfeldt}

\author{Mikhail V. Balabas}

\author{Emil Zeuthen}

\author{Eugene S. Polzik}
\email{polzik@nbi.ku.dk}
\affiliation{Niels Bohr Institute, University of Copenhagen, Copenhagen, Denmark}
	
\begin{abstract}
Continuous measurements of the position of an oscillator become projective on position eigenstates when the measurements are made faster than the coherent evolution.
We evidence an effect of this transition on a spin oscillator within an ensemble of $2\times10^{10}$ room-temperature atoms by observing correlations between the quadratures of the meter light field. 
These correlations squeeze the fluctuations of the light quadratures below the vacuum level. When the measurement is slower than the oscillation, we generate $11.5^{+2.5}_{-1.5}\,\t{dB}$ and detect $8.5^{+0.1}_{-0.1}\,\t{dB}$ of squeezing in a tunable band that is a fraction of the resonance frequency.
When the measurement is as fast as the oscillation, we detect \SI{4.7}{dB} of squeezing that spans more than one decade of frequencies below the resonance.
Our results demonstrate a new regime of continuous quantum measurements on material oscillators, and set a new benchmark for the performance of a linear quantum sensor. 
\end{abstract}

	
\date{\today}
\maketitle
	 

\section{Introduction}

Projective, or von Neumann, measurements collapse the observed quantum system on eigenstates of a Hermitian operator, while more general measurements, described by positive operator-valued measures, collapse the system on states from an overcomplete set~\cite{wiseman_quantum_2009}.
A gradual transition between the two situations can be realized in continuous measurements using meter fields, a canonical example of which is an optical interferometric measurement of the position of a harmonic oscillator~\cite{braginsky_quantum_1992}.
Position measurements are associated with mechanical resonators~\cite{purdy_observation_2013}, collective atomic spins~\cite{hammerer_quantum_2010,grangier_back-action-induced_1994}, ferromagnetic solid-state media~\cite{graf_cavity_2018}, single molecules~\cite{roelli_molecular_2016}, or density waves in liquids~\cite{shkarin_quantum_2019}, that are linearly probed by traveling optical or microwave fields. 
The boundary between generalized and von Neumann measurements occurs at a certain value of the measurement rate~\cite{meng_mechanical_2020}.
When the rate is slower than the oscillation, measurements with the meter in the vacuum input state project the oscillator on coherent states. 
When the rate is faster than the oscillation, measurements project the oscillator on position-squeezed states.  

\begin{figure}[t]
    \centering
    \includegraphics[width=\columnwidth]{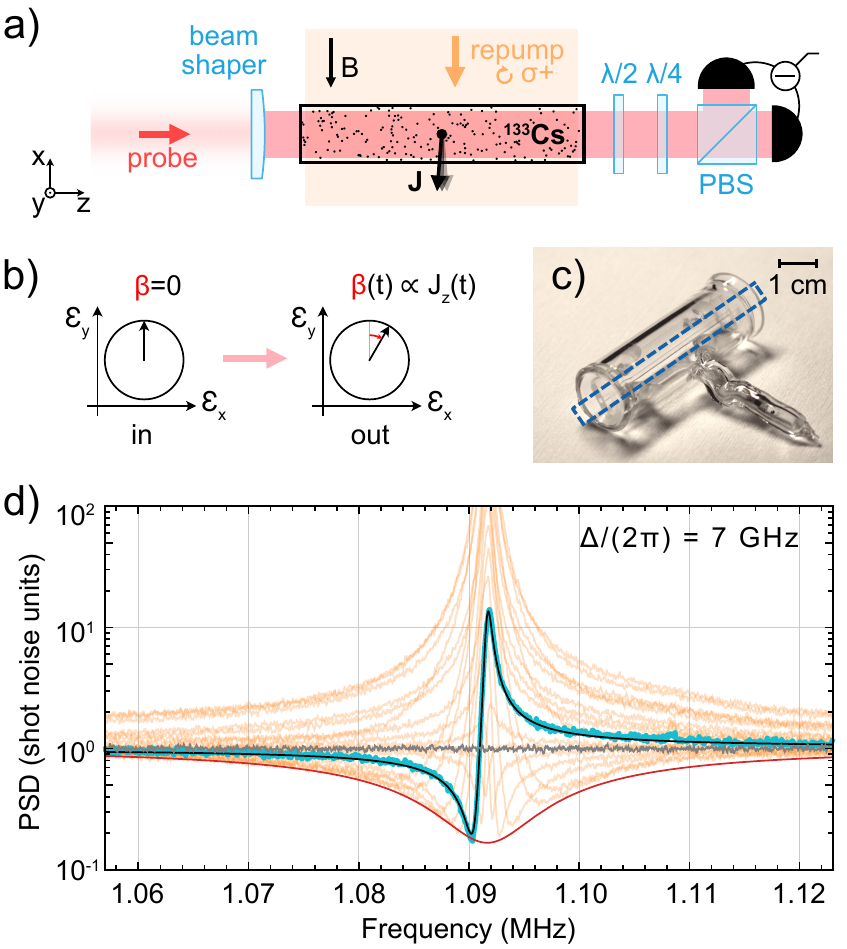}
    \caption{\small{a) An optical probe spatially shaped in a square tophat beam travels through an atomic ensemble with the total spin $\bm{J}$ in a magnetic field $B$, and is detected using balanced polarization homodyning. The detected quadrature is selected using the $\lambda/2$ and $\lambda/4$ waveplates. The total spin is oriented by the repump beam traveling along $x$. PBS: polarization beam splitter. 
    b) The polarization angle $\beta$ of the probe as a meter for the spin projection $\hat{J}_z$. c) A photograph of an anti-spin-relaxation coated cell. The channel with probed atoms is indicated by the blue rectangle. d) The orange curves show power spectral densities (PSD) of homodyne signals recorded at $\Delta/(2\pi) = \SI{7}{GHz}$ at different quadratures. The trace showing the largest squeezing is highlighted by the blue curve. The black curve is the theoretical prediction based on the global fit including all quadratures (see the SI). The gray curve is the shot-noise level. The red curve is the theoretical optimum-quadrature squeezing spectrum.}}
    \label{fig:intro}
\end{figure}

In addition to the oscillator state, the rate of position measurement affects the output state of the meter field~\cite{meng_mechanical_2020}. 
The quadratures of the meter are correlated, and their fluctuations can be below the vacuum level~\cite{fabre_quantum-noise_1994,mancini_quantum_1994}. 
In the slow measurement regime, the correlations and the associated squeezing exist in a narrow frequency band near the resonance, and have a strong frequency dependence due to the time-averaged response of the oscillator to the measurement backaction. 
When the measurement is faster than the oscillation, the correlations and squeezing are broadband and frequency-independent at low frequencies, where the oscillator responds to the backaction instantaneously. 
The detection of squeezing means observing the backaction-driven motion of the oscillator at frequencies much lower than the resonance, which is a necessary condition for position squeezing~\cite{meng_mechanical_2020}. 

The squeezing of the meter light is both a valuable quantum resource and a figure of merit for the purity of the light-oscillator interaction. 
In the slow regime, we realize a measurement of a collective spin of a room-temperature atomic ensemble at a rate fifteen times higher than the rate of thermal decoherence.
The generated squeezing of the meter light reaches $11.5^{+2.5}_{-1.5}\,\t{dB}$ at the output of the cell, exceeding the squeezing demonstrated previously using collective atomic spins~\cite{mccormick_strong_2007,boyer_entangled_2008,thomas_entanglement_2020}, optomechanical cavities~\cite{brooks_non-classical_2012,safavi-naeini_squeezed_2013,purdy_strong_2013}, levitated nanoparticles~\cite{magrini_squeezed_2022,militaru_ponderomotive_2022}, 
and compact on-chip sources utilizing material nonlinearity~\cite{zhang_squeezed_2021}, 
while approaching the results achievable using bulk nonlinear crystals~\cite{vahlbruch_detection_2016}.  
In the fast-measurement regime, we detect broadband squeezing in a bandwidth of several MHz while keeping the backaction-imprecision product~\cite{clerk_introduction_2010} within $\SI{20}{\percent}$ from the value saturating the Heisenberg uncertainty relation. 
These results enable new regimes for sensing surpassing the standard quantum limit~\cite{vyatchanin_quantum_1993,mason_continuous_2019}, 
tests of uncertainty relations for past quantum states~\cite{tsang_optimal_2009,bao_retrodiction_2020},
quantum control of material oscillators~\cite{wilson_measurement-based_2015,rossi_measurement-based_2018,tebbenjohanns_quantum_2021,magrini_real-time_2021},
and links between collective spins and other material systems~\cite{moller_quantum_2017,thomas_entanglement_2020,karg_light-mediated_2020,schmid_coherent_2022}.

\section{Measurements of spin oscillators}

\begin{figure*}[t]
\centering
    \includegraphics[width=\textwidth]{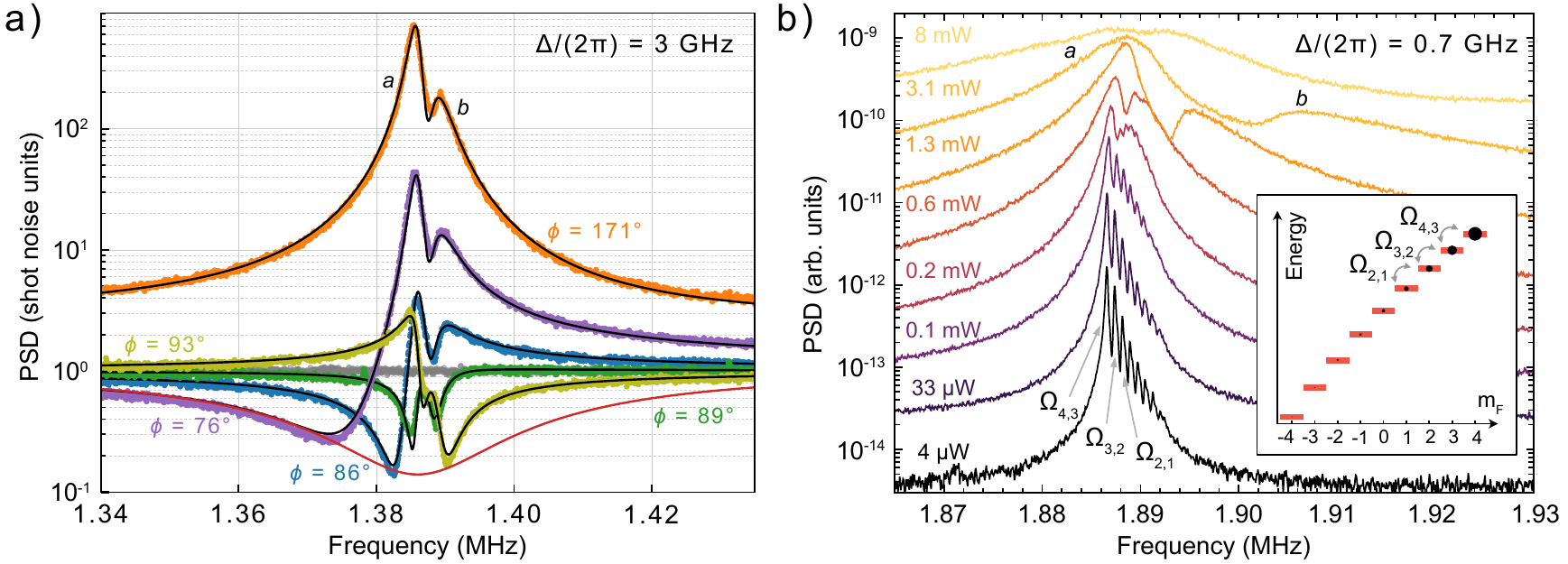}
	\caption{\small{a) Homodyne signal PSDs at $\Delta/(2\pi) = \SI{3}{GHz}$ and different detection angles $\phi$ indicated in the figure. The points are experimental data. The green and orange traces are obtained close to $\hat{P}_\t{L}$ and $\hat{X}_\t{L}$, respectively, and the olive, blue and purple---at intermediate quadratures. The gray points show the shot-noise level. The black curves are theoretical predictions based on the global fit including the spectra at 15 quadratures (see the SI). The red curve is the optimum-quadrature squeezing spectrum predicted by the single-oscillator model.	
	b) The spectra of classically driven motion of the collective spin. 
	The eight peaks visible at low probe powers correspond to bare oscillator modes due to the transitions between adjacent $m_F$ levels. Their frequencies are determined by the linear and quadratic Zeeman energies, and magnitudes are determined by the macroscopic populations of the $m_F$ levels as shown in the inset. 
	The spectra at high powers expose the hybridized oscillator modes. 
	}}
	\label{fig:quadsweep}
\end{figure*}

Linearly polarized light traveling through an oriented atomic medium (as illustrated in \figref{fig:intro}a-b) continuously measures the projection of the total spin on the propagation direction, $\hat{J}_z$, via polarization rotation. This measurement acts back on the spin via quantum fluctuations of ponderomotive torque. When the input light is in a strong coherent state, and the spin satisfies the Holstein-Primakoff approximation~\cite{arecchi_atomic_1972}, the process can be described in terms of linearly coupled pairs of canonically conjugate position and momentum variables. The canonical variables of the spin, $\hat{X}_\t{S}$ and $\hat{P}_\t{S}$, are the normalized projections defined as $\hat{X}_\t{S}=\hat{J}_z/\sqrt{\smash[b]{\hbar\langle J_x\rangle}}$ and $\hat{P}_\t{S}=-\hat{J}_y/\sqrt{\smash[b]{\hbar\langle J_x\rangle}}$, which satisfy the commutation relation $[\hat{X}_\t{S},\hat{P}_\t{S}] = i$. The variables of the light, $\hat{X}_\t{L}$ and $\hat{P}_\t{L}$, are the quadratures proportional to the amplitude and phase differences between the circularly polarized components, respectively. Their commutator is $[\hat{X}_\t{L}(t),\hat{P}_\t{L}(t')]=(i/2) \delta(t-t')$. The Heisenberg uncertainty principle constrains the two-sided spectral densities of the imprecision in the $\hat{P}_\t{L}$-quadrature measurements, $S_\t{imp}$, and the measurement backaction, $S_\t{BA}$, as $\sqrt{S_\t{imp}\, S_\t{BA}}\ge\hbar/2$ (see Ref.~\cite{clerk_introduction_2010} and the SI). This uncertainty relation is saturated if the detection efficiency is perfect and there is no excess measurement noise.

When the ensemble is probed far-detuned from optical transitions, the total spin couples to the probe via the position-measurement Hamiltonian $\hat{H}_\t{int}=-2\hbar\sqrt{\Gamma}\,\hat{X}_\t{L}\hat{X}_\t{S}$, and modifies the probe variables according to the input-output relations~\cite{hammerer_teleportation_2005,thomas_entanglement_2020}
\begin{align}
&\hat{P}_\t{L}^\t{out}(t)=\hat{P}_\t{L}^\t{in}(t)+\sqrt{\Gamma}\, \hat{X}_\t{S}(t),
&\hat{X}_\t{L}^\t{out}(t)=\hat{X}_\t{L}^\t{in}(t),
\end{align} 
where $\Gamma$ is the measurement rate proportional to the optical power. The measurement backaction force is $\hat{F}_\t{QBA}=2\sqrt{\Gamma} \hat{X}_\t{L}^\t{in}$. The response of the spin to the measurement backaction in this situation is described by the Fourier-domain susceptibility $\chi[\Omega]=\Omega_\t{S}/(\Omega_\t{S}^2-\Omega^2-i\Omega\gamma_0)$, where $\Omega_\t{S}$ is the resonance Larmor frequency and $\gamma_0$ is the intrinsic decay rate. 
The response induces correlations between $\hat{X}_\t{L}^\t{out}$ and $\hat{P}_\t{L}^\t{out}$ that can be observed by detecting intermediate quadratures of light, 
$\hat{Q}^\phi_\t{L}=\sin(\phi)\hat{X}_\t{L}^\t{out}+\cos(\phi)\hat{P}_\t{L}^\t{out}$. 
The two-sided spectra of those quadratures, detected by a homodyne with efficiency $\eta$, are given by
\begin{multline}\label{eq:Sphi}
S_\phi[\Omega]=1/4+(\eta\Gamma/2)\,\t{Re}\left(\chi[\Omega]\right)\sin(2\phi)\\
+\eta\Gamma(\Gamma+\gamma_\t{th})|\chi[\Omega]|^2\cos(\phi)^2,
\end{multline}
where $\gamma_\t{th}=(2\,n_\t{th}+1)\gamma_0$ is the thermal decoherence rate. The term $\propto \cos(\phi)^2$ is due to the spin oscillator motion, and the term $\propto \sin(2\theta)$ is due to the cross-correlation between $\hat{X}_\t{S}$ and $\hat{X}_\t{L}^\t{out}$. Negative cross-correlation can squeeze $S_\phi[\Omega]$ below the vacuum level of 1/4.

In a more general situation, the internal dynamics of the collective spin are those of $2F$ harmonic oscillators, where $F$ is the ground-state angular momentum number of the atomic species. Their annihilation operators,
\begin{equation}\label{eq:bmdef}
\hat{b}_m=\frac{1}{\sqrt{\Delta N_m}}\sum_{j=1}^N \ket{m+1}_j\bra{m}_j,
\end{equation} 
are introduced using the multilevel Holstein-Primakoff approximation~\cite{kurucz_multilevel_2010}.
In \eqref{eq:bmdef}, $m$ is the projection quantum number of the single-atom angular momentum on the $x$ axis, $\ket{m+1}_j\bra{m}_j$ are the jump operators between the states $\ket{m}_j$ and $\ket{m+1}_j$ of the individual atoms, and $\Delta N_m=N_{m+1}-N_m$ are the differences in the mean numbers of atoms in the corresponding states. The frequencies of the oscillators are the energy differences between $\ket{m}_j$ and $\ket{m+1}_j$, controlled by an external static magnetic field. The oscillator-light interaction is described by the Hamiltonian
\begin{equation}\label{eq:HamGeneral}
\hat{H}_\mathrm{int} = -2\hbar\sum_{m=-F}^{F-1}\sqrt{\Gamma_{m}} \left(\hat{X}_{m}\hat{X}_\mathrm{L}+\zeta_m\hat{P}_m\hat{P}_\mathrm{L}\right),
\end{equation}
where the quadratures of the modes satisfy $[\hat{X}_{m},\hat{P}_m]=i$, $\Gamma_{m}$ are the measurement rates, and $\zeta_m=\zeta(2m+1)/7$ determine the strengths of dynamical backaction. The common factor $\zeta$ is a function of the optical detuning $\Delta$ and the level structure. 
The deviation of the interaction Hamiltonian~(\ref{eq:HamGeneral}) from that of pure position measurement, $\zeta=0$, results in dynamical-backaction damping with rates $\gamma_{\t{DBA},m}=2\zeta_m\Gamma_m$, and increases the quantum backaction-imprecision product by an amount proportional to $\zeta^2$ (see the SI), which is small in all our experiments. 
The oscillators experience thermal decoherence due to the spontaneous scattering and the collisions of atoms. The thermal occupancy of the intrinsic damping bath is $n_\t{th}=N_m/\Delta N_m$, experimentally found to be independent of $m$.

The multimode structure can affect the response of the spin to the measurement backaction at frequencies close to $\Omega_\t{S}$, while far away from $\Omega_\t{S}$ the spin acts as a single oscillator with $\hat{X}_\t{S}=\sum_m \sqrt{\Gamma_{m}/\Gamma}\hat{X}_{m}$ that is measured at the total rate $\Gamma=\sum_m \Gamma_{m}$ and experiences decoherence at the rate $\gamma_\t{th}=\sum_m \gamma_{\t{th},m}\Gamma_{m}/\Gamma$, where $\gamma_{\t{th},m}$ are the individual decoherence rates of the modes. The quantum cooperativities for the individual modes are defined as the ratios of the measurement and decoherence rates. For the total spin, the cooperativity is $\mathcal{C}_\t{q}=\Gamma/\gamma_\t{th}$.

\section{Experiment}

An ensemble of $N\approx 2\times10^{10}$ cesium-133 atoms at \SI{52}{\degree C} is contained in the \SI{1}{mm}$\times$\SI{1}{mm}$\times$\SI{4}{cm} channel of a glass chip, shown in \figref{fig:intro}c. The channel is coated with paraffin to reduce the spin decoherence from wall collisions~\cite{balabas_polarized_2010}, and is positioned in a homogeneous magnetic field directed along the $x$ axis (\figref{fig:intro}a). The ensemble is continuously probed by a $y-$polarized laser beam propagating in the $z$ direction that has the wavelength \SI{852.3}{nm}, blue-detuned from the $F=4\to F'=5$ transition of the D2 line by $\Delta/(2\pi)=0.7-\SI{7}{GHz}$. The ensemble is also continuously repumped using circularly polarized light resonant with the $F=3\to F'=2$ transition of the D2 line. The combination of spontaneous scattering of probe photons and repumping maintains a steady-state distribution of atoms over the magnetic sublevels of the $F=4$ ground state, which has the macroscopic spin orientation along the magnetic field with polarization $\langle\hat{J}_x\rangle/(NF)\approx0.78$. The steady-state populations are independent of the probe power in our regime, and correspond to the occupancy of the thermal bath $n_\t{th}= 0.9\pm 0.1$. The resonance frequencies of the oscillators are set by the Larmor frequency and split by $0-\SI{40}{kHz}$ in different regimes by the quadratic Zeeman and tensor Stark effects. The Larmor frequency can be positive or negative depending on the orientation of the magnetic field, setting the signs of the effective oscillator masses. We work in the negative-mass configuration \cite{moller_quantum_2017}, but the effects that we observe, in particular the squeezing levels, do not change upon the reversal of the sign of mass (see the SI). The output light is detected using balanced polarization homodyning, which enables shot-noise-limited detection at frequencies down to \SI{10}{kHz}.

\begin{figure*}[t]
\centering
    \includegraphics[width=\textwidth]{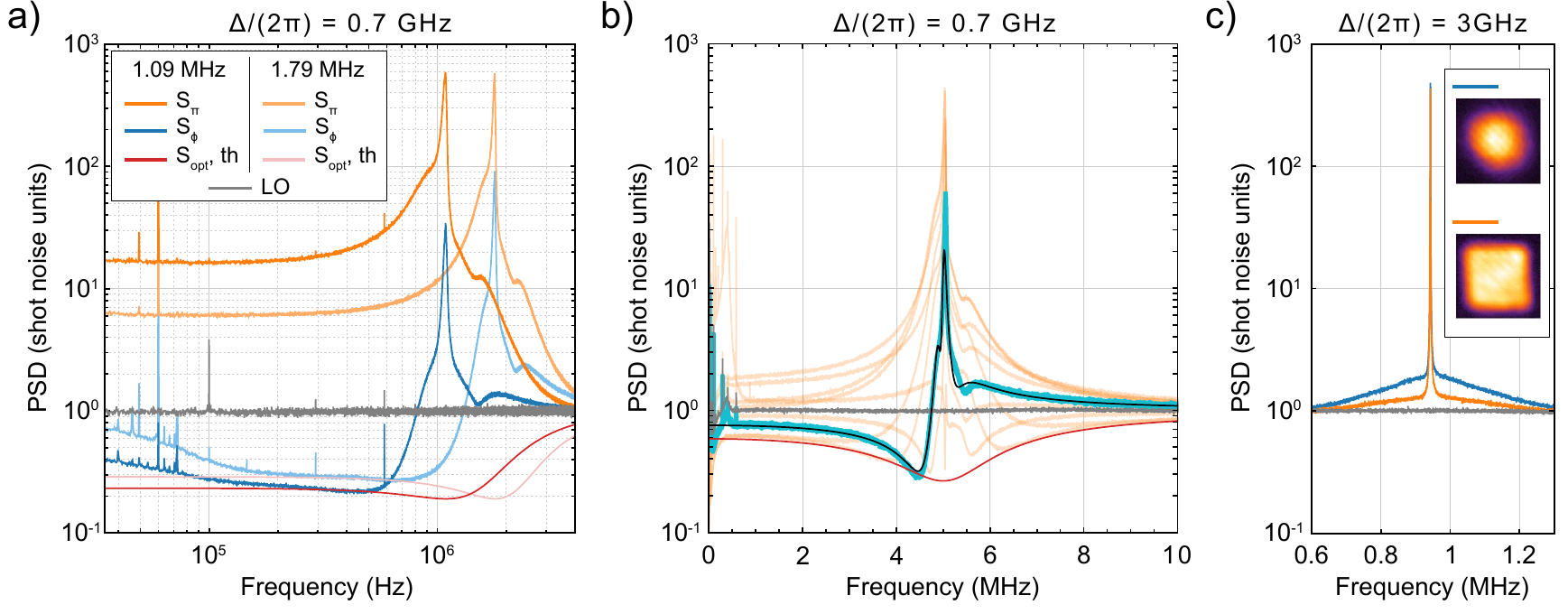}
	\caption{\small{
	a-b) Homodyne signal PSDs at $\Delta/(2\pi)=\SI{0.7}{GHz}$. The gray curves show the experimental shot-noise levels, and the red curves are the theoretical optimum-quadrature squeezing spectra derived from~\eqref{eq:Sphi}.
	a) Spectra for $|\Omega_\t{S}|/(2\pi)=\SI{1.09}{MHz}$ and $\SI{1.79}{MHz}$. 
	The orange and blue curves are measurements with the quadrature angle set to detect $\hat{P}_\t{L}$ and a quadrature $\phi$ close to $\hat{X}_\t{L}$, respectively.
	LO: local oscillator, th: theoretical. 
	b) Orange curves show homodyne spectra recorded at $|\Omega_\t{S}|/(2\pi)=\SI{5}{MHz}$ and at different quadratures $\phi$. The trace with the largest squeezing is highlighted by the blue curve. The black curve is the theoretical prediction based on the global fit including all quadratures (see the SI). 
	c) The spectra taken at the $\hat{P}_\t{L}$ quadrature when the probe beam is Gaussian (blue curve) and tophat (orange curve). The gray curve is the shot noise. The inset shows the beam intensity distributions over the \SI{1}{mm}$\times$\SI{1}{mm} channel cross section recorded without the cell.}} 
	\label{fig:broadband}
\end{figure*}

\section{Results} 

In \figref{fig:intro}d, we present homodyne spectra recorded at the optical detuning $\Delta/(2\pi)=\SI{7}{GHz}$ over a range of detection quadratures $\phi$. In this measurement, dynamical backaction effects are small ($\zeta\approx 0.01$), and the probed spin behaves as a single oscillator subjected to position measurements. The data in \figref{fig:intro}d shows squeezing down to \SI{7.5}{dB}, attained  by the highlighted blue trace. From a global fit of the spectra at all quadratures, we infer the measurement rate $\Gamma/(2\pi)=\SI{13}{kHz}$ and the quantum cooperativity $\mathcal{C}_\t{q}=11$. The measurement rate can be verified directly from \figref{fig:intro}d via the width $\Delta\Omega$ of the frequency band over which squeezing is present in any of the traces, which in the backaction-dominated regime is $\Delta\Omega \sim\Gamma$. The envelope of the traces in \figref{fig:intro}d is described by the spectrum given by \eqref{eq:Sphi} minimized over the detection quadrature at each frequency. Neglecting the imaginary part of the response, the optimum-quadrature spectrum is given by
\begin{equation}\label{eq:Sopt}
S_\t{min}[\Omega]=\frac{1}{4}-\frac{\eta}{2}\frac{\Gamma}{\Gamma+\gamma_\t{th}}\,D\left(\frac{\Omega-\Omega_\t{S}}{\Gamma+\gamma_\t{th}}\right),
\end{equation}
where $D(x)= 1/\left(1+\sqrt{1+4x^2}\right)$. The red curve plotted in \figref{fig:intro}d additionally accounts for 0.7 shot noise units of excess $\hat{P}_\t{L}$-quadrature noise from the thermal motion of fast-decaying spin modes (see \secref{sec:broadModes}). 
This noise is the main limitation for the backaction-imprecision product in this measurement, which equals $1.5\times(\hbar/2)$. 

Due to the scaling $\Gamma\propto 1/\Delta^2$, higher measurement rates are achievable with the probe laser tuned closer to the atomic transition. In \figref{fig:quadsweep}a we present data obtained at the optical detuning of \SI{3}{GHz} using \SI{8.4}{mW} of probe power. In this measurement $\zeta=0.054$, in which case the dynamical backaction results in optical damping and hybridization of the oscillator modes, as well as optical squeezing in the $\hat{X}_\t{L}$-quadrature (see the green trace in \figref{fig:quadsweep}a). Since the thermal decoherence of the oscillators is due to baths at a temperature close to zero, the optical damping improves the maximum magnitude of squeezing by about \SI{0.5}{dB}.
The minimum noise shown by the blue trace in \figref{fig:quadsweep}a is $8.5^{+0.1}_{-0.1}\,\t{dB}$ below the shot noise level. 
The overall detection efficiency of our setup is $\eta=\SI[separate-uncertainty=true]{91\pm 3}{\percent}$, and the transmission loss at the exit window of the cell is \SI{1.6}{\percent}, which means that the magnitude of the squeezing at the exit of the cell is $11.5^{+2.5}_{-1.5}\,\t{dB}$.
The backaction-imprecision product in this measurement is $1.9\times(\hbar/2)$, which is higher than in the measurement at \SI{7}{GHz} detuning due to the higher excess $\hat{P}_\t{L}$-quadrature noise (two shot noise units).



The experimental spectra in \figref{fig:quadsweep}a can be understood as arising from the coupled dynamics of two nearly-degenerate bright modes of the spin, which we refer to as modes $a$ and $b$. 
To extract their effective parameters, we globally fit the set of spectra recorded over an extended range of quadrature angles (see the SI). We find the total measurement rate to be $\Gamma/(2\pi)=\SI{52}{kHz}$, the individual quantum cooperativities to be $\mathcal{C}_\t{q}^a=12$ and $\mathcal{C}_\t{q}^b=4$, and the total cooperativity to be $\mathcal{C}_\t{q}=15$. 
The lower envelope of the experimental traces is in agreement with the optimum-quadrature spectrum predicted by the single-oscillator model using the same $\Gamma$ and $\mathcal{C}_\t{q}$. 

The bright modes $a$ and $b$ emerge due to the coupling of the individual spin oscillators via the common reservoir of the probe optical modes with coupling rates proportional to $\zeta_m$ and $\Gamma_m$. To illustrate this effect, we set the laser detuning to \SI{0.7}{GHz}, where the dynamical backation coefficient is larger, $\zeta=0.18$, and excite the oscillators with classical white noise applied via a magnetic field. 
The spectra of the $P$ quadrature of the output light at different probe powers are shown in \figref{fig:quadsweep}b. 
At the lowest power, the eight bare spin oscillators due to the transitions between adjacent $m_F$ levels are individually resolved.
As the probe power is increased, the resonances first merge in two (the $a$ and $b$ modes) and then three. The macroscopic occupancies of different $m_F$ levels in the atomic ensemble remain the same at all powers, as we separately check, which means that the change in the output spectrum is only due to the coupled dynamics of the collective oscillators.

At the detuning of \SI{0.7}{GHz} from the optical transition, the measurement rate of the spin motion can be as high as the oscillation frequency. 
While around the Larmor resonance, in a frequency band of approximately one hundred kHz, the coupling between individual spin oscillators is pronounced, at frequencies much lower than the resonance the spin behaves as a single oscillator, and the quantum measurement backaction manifests via broadband squeezing of light.
In \figref{fig:broadband}a, we present spectra recorded using \SI{12.8}{mW} of optical probe power at two resonance frequencies, \SI{1.09}{MHz} and \SI{1.79}{MHz}, in which the bandwidth of low-frequency squeezing extends down to \SI{30}{kHz}. 
The minimum noise levels of the homodyne signals (\SI{6.5}{dB} below the shot noise for the \SI{1.09}{MHz} data) are consistent with the quantum cooperativity $\mathcal{C}_\t{q}=8$. 
The measurement rate can be estimated from the signal-to-shot-noise ratio on the $P$ quadrature in \figref{fig:broadband}a using the formula
\begin{equation}
S_{\phi=0}[0]=1/4+\eta \left(\Gamma/\Omega_\t{S}\right)^2,
\end{equation}
which yields $\Gamma/(2\pi)\approx\SI{2}{MHz}$, a value higher than the resonance frequencies. 
To further corroborate the measurement rate, we perform a quadrature sweep with the resonance frequency set to \SI{5}{MHz} and using \SI{10.2}{mW} of probe power (\figref{fig:broadband}b). From fitting this data, we find $\Gamma/(2\pi)=\SI{1.77}{MHz}$, which is consistent within ten percent with the previous estimate corrected for the difference in the probe powers.
Theoretically, the optimum-quadrature noise levels should saturate as the Fourier frequency approaches zero, to a value around $0.22$ shot-noise units for the \SI{1.09}{MHz} data in \figref{fig:broadband}a, while experimental noise levels increase at low frequencies due to excess noise from the atomic ensemble.


The backaction-imprecision product for the measurements in \figref{fig:broadband}a is below $1.2\times(\hbar/2)$ at frequencies higher than \SI{100}{kHz}. This value is closer to saturating the Heisenberg uncertainty relation than the values in the slow-measurement experiments, because the fast-decaying modes are in the backaction-dominated regime, and do not contribute excess thermal noise. The limiting factors for the product in this case are the dynamical backaction and detection inefficiency.

\section{Fast-decaying modes}\label{sec:broadModes}

In addition to the collective oscillators described by the annihilation operators from \eqref{eq:bmdef}, in which all atoms contribute equally, there are other modes of the spin in our system~\cite{shaham_quantum_2020,tang_spin-noise_2020}. The resonance frequencies of these modes coincide with $\Omega_\t{S}$, but their decay rates are limited by the rate of atoms flying through the probe field ($\gamma_{0,\t{flight}}/(2\pi)\approx \SI{300}{kHz}$) rather than collisions with the walls and other atoms ($\gamma_{0,\t{coll}}/(2\pi)\approx\SI{200}{Hz}$). 
The annihilation operators of these modes are  
\begin{equation}
\hat{b}'_{m}=\frac{1}{\sqrt{\Delta N_m\, \langle\Delta g(t)^2\rangle_c}}\sum_{j=1}^N \Delta g_j(t)\, \ket{m}_j\bra{m+1}_j,
\end{equation} 
where $g_j(t)$ are the coupling rates between the optical probe and the individual atoms (see the SI) and $\langle\Delta g^2\rangle_c$ is the squared deviation of the coupling from the mean averaged over classical trajectories, assumed to be the same for all atoms.
The measurement rate of the fast-decaying modes is $\propto  \langle\Delta g^2\rangle_c$, while the measurement rate of the slow-decaying modes is $\propto \langle g\rangle_c^2$. 

An enabling feature of our experiment is the high 3D uniformity of the optical probe field, achieved using a tophat beam configuration, which reduces $\langle\Delta g^2\rangle_c$ and thus the readout of the fast-decaying modes. 
In \figref{fig:broadband}c, we compare the spectra recorded at the $\hat{P}_\t{L}$-quadrature using a tophat and a wide Gaussian probe beam with equal optical powers in the slow-measurement regime. The thermal noise contributed by the fast-decaying modes is reduced from 1 to $0.3$ shot-noise units on resonance upon switching from the Gaussian to the tophat probe. 
The absolute non-uniformity of the coupling~\cite{borregaard_scalable_2016,dideriksen_room-temperature_2021} for the tophat beam is estimated to be $\langle\Delta g^2\rangle_c/\langle g\rangle^2_c=0.6$ based on the camera imaging.

\section{Outlook} 

Continuous measurements that combine high measurement rate, quantum cooperativity, and detection efficiency can be used for single-shot generation of spin-squeezed states and quantum state tomography~\cite{vanner_pulsed_2011}. 
The entanglement link between the material spin and traveling light entailed by the squeezing enables quantum-coherent coupling of spins with other material systems~\cite{thomas_entanglement_2020,karg_light-mediated_2020}. 
While the backaction-imprecision product in all our measurements is already within a factor of two from the Heisenberg bound, it can be further improved by optimizing the probe power for measurements of the $\hat{P}_\t{L}$-quadrature. Our measurements were optimized for quadratures intermediate between $\hat{X}_\t{L}$ and $\hat{P}_\t{L}$ (i.e. for ``variational" readout~\cite{vyatchanin_quantum_1993}) which can yield superior results~\cite{habibi_quantum_2016} in quantum sensing and control.

This work also establishes room-temperature atomic spin oscillators as a practical platform for engineering quantum light with high levels of squeezing, which is a basic resource for interferometric sensing and optical quantum information processing~\cite{zhang_squeezed_2021}. 
The highest demonstrated squeezing, reaching \SI{8.5}{dB} at the detection, is narrowband, but its frequency can be tuned by the magnetic field without degrading the level within the range of approximately $0.8-\SI{5}{MHz}$ in our experiments.

\section{Acknowledgements}
The authors thank Micha\l{} Parniak, Jörg Müller, Rebecca Schmieg, and Ivan Galinskiy for general help and useful discussions. 
This work was supported by the European Research Council (ERC) under the Horizon 2020 (grant agreement No 787520) and by VILLUM FONDEN under a Villum Investigator Grant no. 25880.
SF acknowledges funding from the European Union’s Horizon 2020 research program under the Marie Sklodowska-Curie grant agreement No. 847523 ``INTERACTIONS".

\bibliography{references}

\end{document}


\appendix

\setcounter{page}{1}
\renewcommand{\thepage}{SI~\arabic{page}}

\setcounter{figure}{0}
\renewcommand{\thefigure}{SI\arabic{figure}}

\setcounter{table}{0}
\renewcommand{\thetable}{SI\arabic{table}}

\setcounter{equation}{0}
\renewcommand{\theequation}{SI~\thesection.\arabic{equation}}

\onecolumngrid
\begin{center}
\textbf{{\Large Supplementary Information}}
\end{center}

\title{Squeezed light from an oscillator measured at the rate of oscillation}

\author{Christian Bærentsen}

\author{Sergey A. Fedorov}
\email{sergey.fedorov@nbi.ku.dk}

\author{Christoffer Østfeldt}

\author{Mikhail V. Balabas}

\author{Emil Zeuthen}

\author{Eugene S. Polzik}
\email{polzik@nbi.ku.dk}

\affiliation{Niels Bohr Institute, University of Copenhagen, Copenhagen, Denmark}
\maketitle
\onecolumngrid

\tableofcontents

\section{Experimental setup}
\label{app:exp}

\begin{figure}[h]
	\includegraphics[width=0.6\textwidth]{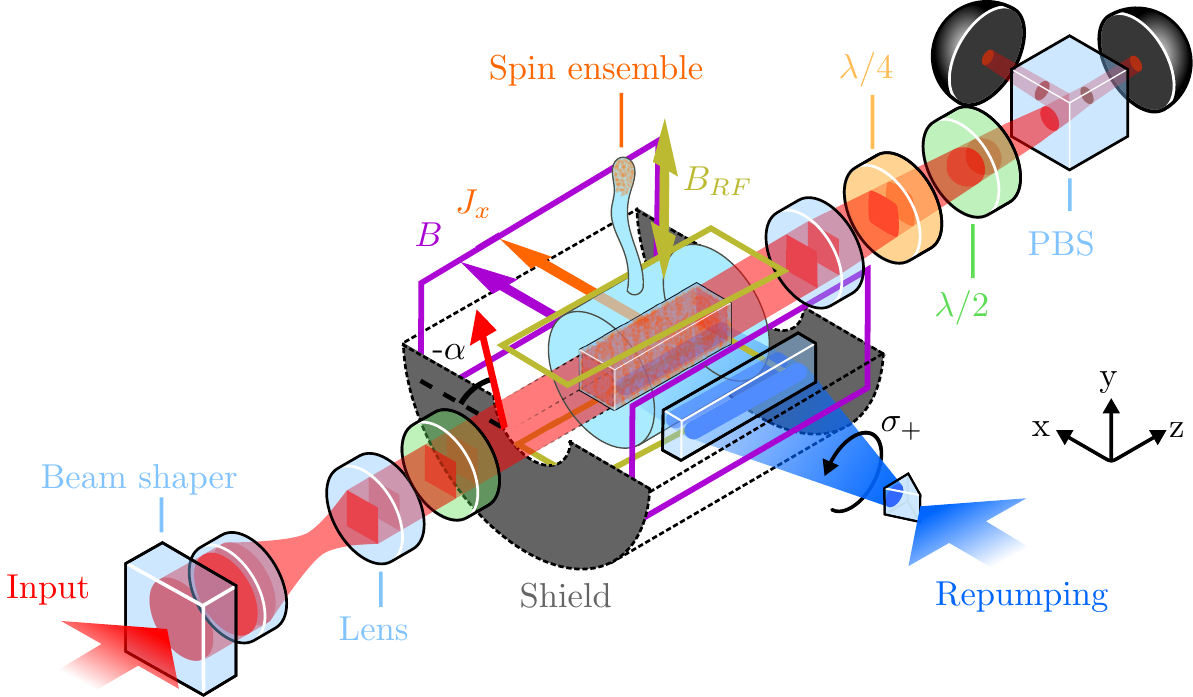}
	\caption{Experimental setup. A linearly polarized light probe is spatially shaped as a square-tophat beam. The probe interacts with an optically polarized ensemble of Cesium atoms located in a glass chip. The macroscopic atomic polarization $J_x$ is oriented along the magnetic field $B$. The optical probe and the atomic ensemble interact via Faraday interaction in the dispersive regime. The output probe light is detected using a polarization self-homodyning setup. PBS: Polarizing beamsplitter. $\lambda/2$: Half wave plate. $\lambda/4$: Quarter wave plate. Beam shaper: Gaussian-to-tophat beam-shaping lens.}
	\label{fig:si:setup}
\end{figure}

A detailed schematic of our experimental setup is presented in \figref{fig:si:setup}.
The probed cesium-133 atoms are located in a channel of a glass chip with \SI{1}{mm}$\times$\SI{1}{mm} cross-section and \SI{40}{mm} length. The chip is enclosed in a glass cell, which has a stem attached to it that contains a piece of cesium metal providing a reservoir of atoms. The cell interior is coated with an anti-spin-relaxation paraffin coating to decrease the decoherence due to the collisions of atoms with walls.
The cell is heated to $\SI[separate-uncertainty=true]{52\pm2}{^{\circ}C}$ and placed in a stationary homogeneous magnetic field directed along the $x$ axis, which is created by a pair of rectangular coils parallel to the $yz$ plane. Additional time-dependent magnetic field directed along the $y$ axis can be created using another pair of coils parallel to the $xz$ plane, which has the effect of applying a classical force to the atomic oscillator. The cell and the entire set of coils are enclosed in a multi-layer magnetic shield, including $\mu$-metal layers to eliminate the magnetic field of the Earth and an aluminum layer to protect the spins from external high-frequency magnetic noise.

The atoms interact with two light beams: the probe, which is linearly polarized and propagates along the channel, and the repump, which is circularly polarized and propagates perpendicular to the channel, along the $x$ axis. Both light beams have wavelengths around \SI{852.3}{nm}, close to the D2 transition from the ground state of Cs.
The ground state of Cs is split into two hyperfine levels, with the magnetic momentum numbers $F=3$ and $F=4$, and each hyperfine level is further split into $(2F+1)$ magnetic sublevels.
The repump beam is produced by a diode laser and has the power in the range of $8-\SI{10}{mW}$.
It is blue-detuned by \SI{80}{MHz} from the $F=3\to F'=2$ transition of the D2 line, and resonant with all transitions $F=3\to F'=2,3,4$ within the Doppler linewidth, where the primes denote electronically excited states.
The cross-section of the chip channel containing atoms is chosen to be square to avoid lensing of the repump beam. In order to uniformly illuminate the elongated channel, the repump beam is shaped by a combination of a Powell lens and a cylindrical collimating lens. The repump transfers all atoms to $F=4$ level, and simultaneously creates macroscopic spin orientation in the ensemble because of its circular polarization. The chirality of the polarization, $\sigma_{+}$ or  $\sigma_{-}$, determines the sign of the mass of the oscillator~\cite{moller_quantum_2017}. Our experiments are done with a negative-mass oscillator, but the results, including the observed levels of squeezing, are largely independent of the sign of the mass (see \secref{sec:si:massFlip}).
The probe beam is blue-detuned by $0.7-\SI{7}{GHz}$ from $F=4\to F'=5$ transition; it is produced by a Ti:Sa laser and has the power up to \SI{13}{mW}. The probe interacts with the ensemble in the dispersive regime, but the residual spontaneous scattering of photons from it contributes to the spin decoherence.
The linear polarization of the probe is set along the $y$ axis to maximize the optical damping by the dynamical backaction (which nevertheless remains small), and simultaneously decouple the spin from the classical intensity fluctuations. The small amount of optical damping in our experiments improves the maximum observed level of squeezing (see \secref{sec:si:optDamping}).
The decoherence rate due to the spontaneous scattering is proportional to the probe power, and is the primary limitation for the achievable quantum cooperativity in our work.
The distribution of the atoms among the magnetic sublevels is determined by the interplay of the spontaneous scattering processes due to the probe and the repump beam, and is independent of the probe power and detuning within our range of parameters.

After the interaction with the atomic ensemble, the relevant quadratures of the probe beam are detected using polarization homodyning. The quadrature angle is selected using a combination of a quarter waveplate and a half waveplate. A key advantage of the polarization homodyning method is the perfect spatial overlap between the detected modes of light and the local oscillator. The electronic noise floor of the photodetector is typically about \SI{30}{dB} below the shot noise level and hence is negligible.

The maximum narrowband squeezing of light observed in the regime when $\Gamma\ll|\Omega_S|$ is approximately independent of the Larmor frequency within the range of Larmor frequencies between \SI{0.8}{MHz} and \SI{5}{MHz}. At low frequencies, the limitation is due to classical noises acting on the spins, and at high frequencies due to the inhomogeneity of the magnetic field, which could be straightforwardly improved.

In order to minimize the coupling to the fast-decaying modes of the spin ensemble (see \secref{sec:si:spinOscillators}), the probe beam is shaped into a square tophat beam using a high-transmission beam shaping lens (Topag GTH-3.6-1.75FA), and an additional system of regular spherical lenses described in \secref{sec:si:tophat}. The resulting beam has a supergaussian intensity cross section $I(x,y)\propto \exp(-2(x/w_x)^{2n}-2(y/w_y)^{2n})$ with $n\approx 3.2$ and $2w_x\approx2w_y\approx\SI{0.84}{mm}$, which change negligibly in the $z$ direction over the length of the cell channel.
The on-resonance extraneous thermal noise in the slow-measurement regime was experimentally found to be lower by a factor of 3.6 for the tophat beam probe compared to the Gaussian beam probe with the maximum width allowed by the cell channel.
The transmission of the probe beam through the cell reaches 96.8$\%$, limited by the reflection and scattering of light upon hitting the cell windows, with the loss of light due to the clipping of the beam being negligible.
In order to infer the generated level of squeezing from detected, we assume that the transmission loss is equally contributed by the input and the output windows.

\section{The modes of an ensemble of moving atoms interacting with light}\label{sec:si:spinOscillators}
In this section, we describe $N$ moving atoms interacting with the probe light field, and derive input-output relations for the optical quadratures in terms of two types of collective spin oscillator modes: usual Larmor precession modes, and modes scrambled by the atomic motion. 
Individual atoms interact with the light field with the strengths $g_k(t)$ (where $k=1,..,N$ is the integer index that labels the atoms) that is proportional to the intensity of the light field at their instantaneous position.
The interaction strengths randomly change in time as atoms move inside the cell. The motions of different atoms are assumed to have the same statistical properties and be uncorrelated between each other. 
The statistics of motion are characterized by decomposing the couplings into their mean value, $\bar{g}$, and deviations, $\Delta g_k(t)$, 
\begin{equation}
g_k(t)=\bar{g}+\Delta g_k(t),
\end{equation}
and specifying the motional correlation function, $R(\tau)$, 
\begin{equation}
\frac{\langle \Delta g_k(t_1)\,\Delta g_l(t_2)\rangle_{c}}{\langle \Delta g(t)^2\rangle_{c}}=\delta_{kl}R(t_1-t_2),
\end{equation}
where $\delta_{kl}$ is the Kronecker symbol and $\langle \cdot\rangle_{c}$ denotes motional averaging (following the notation of Ref.~\cite{shaham_quantum_2020}, to separate from the quantum averaging $\langle \cdot\rangle$). The normalization factor, $\langle\Delta g(t)^2\rangle_{c}$, is the mean squared deviation among the individual atom-light couplings. According to the ergodic hypothesis, the result of the averaging is the same regardless of whether it is done over the time or the realizations of the ensemble.

The dispersive interaction between the light and the $k-$th atom in the ensemble is described by the Hamiltonian~\cite{sherson_deterministic_2007,thomas_entanglement_2020}
\begin{equation}
\hat{H}_\t{int}^{(k)}=\hbar\, g_k(t)\left[a_0 \hat{I}+a_1 \hat{S}_z \hat{j}_z^{(k)}+a_2\left(\hat{I}\, \hat{j}_z^{(k)}\hat{j}_z^{(k)}-2\hat{S}_x\left(\hat{j}_x^{(k)}\hat{j}_x^{(k)}-\hat{j}_y^{(k)}\hat{j}_y^{(k)}\right)-2\hat{S}_y\left(\hat{j}_x^{(k)}\hat{j}_y^{(k)}+\hat{j}_y^{(k)}\hat{j}_x^{(k)}\right)\right)\right],
\end{equation}
where $\hat{S}_{x,y,z}$ are the Stokes parameters of the input light~\cite{sherson_deterministic_2007}, $\hat{I}$ is the intensity of the input light, and the parameters $a_{0,1,2}$ are functions of the level structure and the laser detuning from the optical transition~\cite{vasilyev_quantum_2012}. After linearization assuming a strong coherent $y$-polarized light probe with the mean amplitude $\bar{a}$, the Hamiltonian is expressed as
\begin{equation}\label{eq:si:Hintj}
\hat{H}^{(k)}_\t{int}= \hat{H}^{(k)}_\t{Stark}-\hbar\frac{\bar{a}g_k(t)}{\sqrt{2}}\left[a_1 \hat{j}_z^{(k)}\hat{X}_\t{L}-2a_2\left(\hat{j}_x^{(k)}\hat{j}_y^{(k)}+\hat{j}_y^{(k)}\hat{j}_x^{(k)}\right)\hat{P}_\t{L}\right],
\end{equation}
where the Stark Hamiltonian $\hat{H}^{(k)}_\t{Stark}=\hbar g_k(t)\left[a_0 + a_2\left(\hat{j}_x^{(k)}\hat{j}_x^{(k)}-\hat{j}_y^{(k)}\hat{j}_y^{(k)}+\hat{j}_z^{(k)}\hat{j}_z^{(k)}\right)\right]\hat{I}$ describes the energy shifts due to the dynamic Stark effect,
and $\hat{X}_\t{L}$ and $\hat{P}_\t{L}$ are the polarization quadratures of the light field normalized such that they satisfy the commutation relation
\begin{equation}
[\hat{X}_\t{L}(t_1),\hat{P}_\t{L}(t_2)]=(i/2)\delta(t_1-t_2).
\end{equation}
The spin components of individual atoms $\hat{j}^{(k)}_{x,y,z}$ can be expressed in terms of the jump operators $\hat{\sigma}_{n,m}^{(k)}$ between the ground state sublevels,
\begin{equation}
\hat{\sigma}_{n,m}^{(k)}= \ket{n}_k\bra{m}_k,
\end{equation}
where $m,n=-F,...,F$ is the projection of the angular momentum on the $x$ axis (which coincides with the direction of the magnetic field), and $F$ is the total angular momentum quantum number of the ground state level.
In this notation,
\begin{equation}\label{eq:si:Hintsigma}
H^{(k)}_\t{int}=\hat{H}^{(k)}_\t{Stark}+\hbar\frac{\bar{a}g_k(t)}{2\sqrt{2}}\sum_{m=-F}^{F-1}C_m\left( ia_1\left(\hat{\sigma}_{m+1,m}^{(k)}-\hat{\sigma}_{m,m+1}^{(k)}\right)\hat{X}_\t{L}+2(2m+1)a_2\left(\hat{\sigma}_{m+1,m}^{(k)}+\hat{\sigma}_{m,m+1}^{(k)}\right)\hat{P}_\t{L}\right),
\end{equation}
where $\hat{H}^{(k)}_\t{Stark}=\hbar\sum_{m}g_k(t)\left(a_0+a_2m^2\right)\hat{I}\,\hat{\sigma}_{m,m}^{(k)}$ is the Stark energy, and $C_m=\sqrt{F(F+1)-m(m+1)}$ are Clebsch–Gordan coefficients. When transiting from \eqref{eq:si:Hintj} to \eqref{eq:si:Hintsigma} we neglected the terms involving second-order coherences that only couple to $\hat{I}$ and are negligibly small in our case.

The individual atomic spins are precessing in a homogeneous magnetic field directed along the $x$ axis.
Taking the zero of the energy scale to be the ground state energy of free atoms, the Hamiltonian of the precession is expressed as
\begin{equation}
\hat{H}^{(k)}_\t{S}=\sum_{m=-F}^{F} E_{\t{Zeem},m}\,\hat{\sigma}_{m,m}^{(k)},
\end{equation}
where $E_{\t{Zeem},m}$ are the Zeeman energies of the magnetic sublevels that include contributions linear and quadratic in $m$. The total Hamiltonian of all atoms,
\begin{equation}
\hat{H}=\sum_{k=1}^N \left(\hat{H}^{(k)}_\t{S}+\hat{H}^{(k)}_\t{int}\right),
\end{equation}
can be expressed using collective operators: the total numbers of atoms in the magnetic sublevels, denoted by $\hat{N}_m$, and two sets of coherences between neighboring $m$ levels, denoted by $\hat{\Sigma}_{m}$ and $\hat{\Sigma}_{m}'$. The operators are defined as
\begin{align}
\hat{N}_m=\sum_{k=1}^N \hat{\sigma}_{m,m}^{(k)},&
&\hat{\Sigma}_m=\sum_{k=1}^N \hat{\sigma}_{m+1,m}^{(k)},&
&\hat{\Sigma}'_m=\frac{1}{\sqrt{\langle\Delta g^2\rangle_{c}}}\sum_{k=1}^N \Delta g_k(t)\, \hat{\sigma}_{m+1,m}^{(k)},
\end{align}
where $m=-F,...,F-1$ for the $\Sigma$ operators and $m=-F,...,F$ for the $N$ operators.
The expression for the Hamiltonian, neglecting a small contribution due to the inhomogeneity of the Stark shift, is
\begin{multline}\label{eq:si:HtotSigma}
\hat{H}=\sum_{m=-F}^{F} E_{m}\,\hat{N}_m+\hbar\sum_{m=-F}^{F-1}\frac{\bar{g}\bar{a}a_1}{2\sqrt{2}}C_m\left( i\left(\hat{\Sigma}_{m}-\hat{\Sigma}_{m}^\dagger\right)\hat{X}_\t{L}+\zeta_m\left(\hat{\Sigma}_{m}+\hat{\Sigma}_{m}^\dagger\right)\hat{P}_\t{L}\right)\\
+\hbar\sum_{m=-F}^{F-1}\frac{\sqrt{\langle\Delta g^2\rangle_{c}}\,\bar{a}a_1}{2\sqrt{2}}C_m\left( i\left(\hat{\Sigma}'_{m}-\hat{\Sigma}_{m}^{\prime\,\dagger}\right)\hat{X}_\t{L}+\zeta_m\left(\hat{\Sigma}'_{m}+\hat{\Sigma}_{m}^{\prime\,\dagger}\right)\hat{P}_\t{L}\right),
\end{multline}
where $\zeta_m=2(2m+1)a_2/a_1$, and $E_{m}=E_{\t{Zeem},m}+E_{\t{Stark},m}$ is the sum of the Zeeman and the Stark energies. In the limit of a large number of atoms in the ensemble, the two sets of $\hat{\Sigma}_{m}$ operators are independent and have constant commutators,
\begin{align}
&\left[\hat{\Sigma}_n,\hat{\Sigma}_m^\dagger\right]=\delta_{nm}\left(\hat{N}_{m+1}-\hat{N}_m\right)&&\underset{N\gg 1}{\longrightarrow}&&\delta_{nm}\left(N_{m+1}-N_m\right),\\
&\left[\hat{\Sigma}_n,\hat{\Sigma}_{m}^{\prime\,\dagger}\right]=\delta_{nm}\sum_k\frac{\Delta g_k(t)}{\sqrt{\langle\Delta g^2\rangle_{c}}}\left(\hat{\sigma}_{m+1,m+1}^{(k)}- \hat{\sigma}_{m,m}^{(k)}\right)&&\underset{N\gg 1}{\longrightarrow}&& 0,\\
&\left[\hat{\Sigma}'_n,\hat{\Sigma}_{m}^{\prime\,\dagger}\right]=\delta_{nm}\sum_j\frac{\Delta g_j(t)^2}{\langle\Delta g^2\rangle_{c}}\left(\hat{\sigma}_{m+1,m+1}^{(j)}- \hat{\sigma}_{m,m}^{(j)}\right)&&\underset{N\gg 1}{\longrightarrow}&&\delta_{nm}\left(N_{m+1}-N_m\right),
\end{align}
where $m,n=-F,...,F-1$, and $N_m=\langle \hat{N} \rangle$ are the average macroscopic populations of the magnetic sublevels.
By normalizing the $\Sigma$ operators to satisfy the canonic commutation relations, we can introduce two sets of bosonic modes, $\hat{b}_m$ and $\hat{b}'_m$, that appear in the main text,
\begin{align}
&\hat{b}_m=\hat{\Sigma}_m/\sqrt{\Delta N_m},
&\hat{b}'_m=\hat{\Sigma}'_m/\sqrt{\Delta N_m},
\end{align}
where $\Delta N_m=N_{m+1}-N_m$.
The modes described by $\hat{b}_m$ are those usually identified with the Larmor precession of the spin ensemble as a whole. They experience coupling to the probe light that is averaged over the atomic trajectories~\cite{borregaard_scalable_2016}, and their coherence time is high, limited by the reorientation of individual spins due to the collisions with the walls and between each other, and by the spontaneous scattering of probe photons. The modes described by $\hat{b}'_m$ experience additional damping and decoherence due to the atoms flying in and out of the probe beam. We refer to them as the fast-decaying modes. Introducing the quadratures of the spin oscillators,
\begin{align}\label{eq:atomicOscQuad}
&\hat{X}_m\equiv\frac{1}{i\sqrt{2}}\left(\hat{b}_{m}-\hat{b}_{m}^\dagger\right),&
&\hat{P}_m\equiv-\frac{1}{\sqrt{2}}\left(\hat{b}_{m}+\hat{b}_{m}^\dagger\right),&
&\hat{X}'_m\equiv\frac{1}{i\sqrt{2}}\left(\hat{b}'_{m}-\hat{b}_{m}^{\prime\,\dagger}\right),&
&\hat{P}'_m\equiv-\frac{1}{\sqrt{2}}\left(\hat{b}_{m}'+\hat{b}_{m}^{\prime\,\dagger}\right),
\end{align}
which satisfy $[\hat{X}_m,\hat{P}_m]=i$ and $[\hat{X}'_m,\hat{P}'_m]=i$, and using the fact that, in the Holstein-Primakoff approximation, the numbers of atoms in the $m-$th levels satisfy
\begin{equation}
\hat{N}_m\approx N_m+\frac{1}{2}\left(\hat{b}^\dagger_m\hat{b}_m+\hat{b}_m^{\prime\,\dagger}\hat{b}'_m-\hat{b}^\dagger_{m-1}\hat{b}_{m-1}-\hat{b}_{m-1}^{\prime\,\dagger}\hat{b}'_{m-1}+\t{h.c.}\right),
\end{equation}
the total Hamiltonian in \eqref{eq:si:HtotSigma} is expressed as
\begin{multline}\label{eq:si:hamOscFull}
\hat{H}=\hbar\sum_{m=-F}^{F-1}\left[\frac{\Omega_m}{2}\left(\hat{X}_m^2+\hat{P}_m^2\right)+\frac{\Omega_m}{2}\left(\hat{X}_m^{\prime\,2}
+\hat{P}_m^{\prime\,2}\right)\right.\\\left.-2\sqrt{\Gamma_m}\left(\hat{X}_m\hat{X}_\t{L}+\zeta_m\hat{P}_m\hat{P}_\t{L}\right)
-2\sqrt{\Gamma'_m}\left(\hat{X}'_m\hat{X}_\t{L}+\zeta_m\hat{P}'_m\hat{P}_\t{L}\right)\right],
\end{multline}
which is a Hamiltonian of $4F$ oscillators linearly coupled to a propagating field. The frequencies $\Omega_m$ are determined by the energy splittings between different magnetic sublevels due to the Zeeman and Stark effects,
\begin{equation}
\hbar\Omega_m=E_{\t{Zeem},m}-E_{\t{Zeem},m+1}-\hbar\bar{g}a_2I(2m+1),
\end{equation}
and
the measurement rates for the slow- and the fast-decaying modes are identified as
\begin{align}
&\Gamma_m=\bar{g}^2(\bar{a}a_1\,C_m)^2\Delta N_m/16,&
&\Gamma'_m=\langle\Delta g^2\rangle_{c}\,(\bar{a}a_1\,C_m)^2\Delta N_m/16.
\end{align}
The input-output relations for the quadratures of the light field are derived based on \eqref{eq:si:hamOscFull} as described in Ref.~\cite{hammerer_teleportation_2005}. They are given by
\begin{align}
&\hat{X}_\t{L}^\t{out}(t)=\hat{X}_\t{L}^\t{in}(t)-\sum_{m=-F}^{F-1}\zeta_m\left(\sqrt{\Gamma_m}\hat{P}_{m}(t)+\sqrt{\Gamma'_{m}}\hat{P}'_{m}(t)\right),\label{eq:si:XI-O}\\
&\hat{P}_\t{L}^\t{out}(t)=\hat{P}_\t{L}^\t{in}(t)+\sum_{m=-F}^{F-1}\left(\sqrt{\Gamma_m}\hat{X}_{m}(t)+\sqrt{\Gamma'_{m}}\hat{X}'_{m}(t)\right),\label{eq:si:PI-O}
\end{align}
and the Heisenberg equations of motion for the slow-decaying modes are
\begin{align}
&\frac{d}{dt}\hat{X}_m(t)=\Omega_m \hat{P}_m(t)-\sum_{n=-F}^{F-1}\zeta_m\sqrt{\Gamma_m}\left(\sqrt{\Gamma_n}\hat{X}_{n}(t)+\sqrt{\Gamma'_{n}}\hat{X}'_{n}(t)\right)-2\zeta_m\sqrt{\Gamma_m}\hat{P}_\t{L}^\t{in}(t),\label{eq:si:XHhc}\\
&\frac{d}{dt}\hat{P}_m(t)=-\Omega_m \hat{X}_m(t)-\sum_{n=-F}^{F-1}\zeta_n\sqrt{\Gamma_m}\left(\sqrt{\Gamma_n}\hat{P}_{n}(t)+\sqrt{\Gamma'_{n}}\hat{P}'_{n}(t)\right)+2\sqrt{\Gamma_m}\hat{X}_\t{L}^\t{in}(t).\label{eq:si:PHhc}
\end{align}
Eq.~(\ref{eq:si:XHhc}-\ref{eq:si:PHhc}) show that the oscillators experience damping or antidamping by dynamical backaction with the rates $\gamma_{\t{DBA},m}=2\zeta_m\Gamma_m$, and are coupled between each other at the rates $\sqrt{\gamma_{\t{DBA},m}\gamma_{\t{DBA},n}}$ due to the interaction with the common optical bath. For practical calculations, intrinsic dissipation due to the atomic collisions and spontaneous scattering is added to Eq.~(\ref{eq:si:XHhc}-\ref{eq:si:PHhc}) using the usual quantum Langevin approach~\cite{thomas_entanglement_2020}. The temperatures of the effective thermal baths can be determined from the equilibrium numbers of excitation in the modes in the absence of probing, $n_\t{th}\equiv\langle \hat{b}^\dagger_m\hat{b}_m\rangle=(N_m/\Delta N_m)$, which are calculated directly from the definitions of $\hat{b}_m$ under the assumption that the processes that determine the equilibrium populations $N_m$ affect all atoms independently.

The Heisenberg equations of motion describing the evolution of the modes from the fast-decaying family are identical to Eq.~(\ref{eq:si:XHhc}-\ref{eq:si:PHhc}), except that they include additional terms due to the explicit time dependence of their operators. These terms are more convenient to present for the annihilation operators than for the quadratures, they are given by
\begin{equation}
\frac{d}{dt}\hat{b}'_m(t)=-i\left[\hat{b}'_m,\hat{H}\right]+\frac{1}{\sqrt{\Delta N_m\,\langle\Delta g^2\rangle_{c}}}\sum_{k=1}^N \left(\frac{d}{dt}\Delta g_k(t)\right) \hat{\sigma}_{m+1,m}^{(k)},
\end{equation}
where $-i[\hat{b}'_m,\hat{H}]$ contributes the terms due to the coherent evolution and the coupling to the light field that are completely analogous to those present in Eq.~(\ref{eq:si:XHhc}-\ref{eq:si:PHhc}). The added terms give rise to both extra dissipation and fluctuations. If the motional correlation function is exponential, $\langle \Delta g_k(t_1)\,\Delta g_k(t_2) \rangle\propto e^{-\gamma_b|t_1-t_2|/2}$, as it was suggested in \cite{borregaard_scalable_2016}, the stochastic evolution of $\Delta g_k(t)$ can be modeled by the Ornstein–Uhlenbeck process,
\begin{equation}
\frac{d}{dt}\Delta g_k(t)=-\frac{\gamma_{b}}{2}\Delta g_k(t)+\sqrt{\gamma_{b}} f_k(t),
\end{equation}
where $\langle f_k(t_1) f_k(t_2)\rangle_{c}=\langle\Delta g^2\rangle_{c}\,\delta(t_1-t_2)$.
In this case, the extra terms in the Heisenberg-Langevin equations for $\hat{b}'$ can be re-expressed as
\begin{equation}
\frac{d}{dt}\hat{b}'_m(t)=-i\left[\hat{b}'_m,\hat{H}\right]-\frac{\gamma_{b}}{2}\hat{b}'_m(t)+\sqrt{\gamma_{b}}\hat{\mathcal{F}}'_b(t),
\end{equation}
where $\langle \hat{\mathcal{F}}^{\prime\dagger}_b(t_1) \hat{\mathcal{F}}'_b(t_2)\rangle=n_\t{th}\delta(t_1-t_2)$ and $n_\t{th}=N_m/\Delta N_m$ is the thermal occupancy of the bath. While the atomic motion increases the decoherence rate, the thermal bath occupancies for the fast- and slow-decaying modes are the same.

\section{The backaction-imprecision product in homodyne detection}

The two conjugated quadratures of the probe light that after interaction with the atomic ensemble, $\hat{X}^\t{out}_\t{L}$ and $\hat{P}^\t{out}_\t{L}$, as well as any intermediate quadrature $\hat{Q}^\phi_\t{L}$,
\begin{equation}
\hat{Q}^\phi_\t{L}(t)=\sin(\phi)\hat{X}_\t{L}^\t{out}(t)+\cos(\phi)\hat{P}_\t{L}^\t{out}(t),
\end{equation}
can be detected by balanced polarization homodyning after passing the output light through a combination of a half and a quarter waveplates. The rotation angles of the waveplates allow setting the detection angle $\phi$. The two-sided power spectral density (PSD) of the photocurrent signal is given by
\begin{equation}
S_\phi[\Omega]=\frac{1}{4}\left(1-\eta\right) +\eta\int_{-\infty}^{\infty} e^{i\Omega \tau}\left\langle\hat{Q}^\phi_\t{L}(t+\tau)\,\hat{Q}^\phi_\t{L}(t) \right\rangle d\tau,
\end{equation}
where $\eta$ is the detection efficiency.
When the optical field is in the vacuum state, its correlation is given by $\langle\hat{Q}^\phi_\t{L}(t+\tau)\,\hat{Q}^\phi_\t{L}(t) \rangle=(1/4)\delta(\tau)$, and therefore $S_\phi[\Omega]=1/4$; this value is the shot noise level. The observation $S_\phi[\Omega]<1/4$ means that some of the Fourier-domain modes of light are in squeezed states.

The spectral density of the photocurrent when the homodyne is tuned to detect the $P$ quadrature is given by
\begin{equation}
S_\phi[\Omega] =\frac{1}{4} +\eta\Gamma\, S_{X_S X_S}[\Omega]+\eta\,S_{PP,\t{ext}}[\Omega],
\end{equation}
where $S_{X_S X_S}[\Omega]$ is the spectrum of the total spin motion, and $S_{PP,\t{ext}}[\Omega]$ is the extraneous noise. In the slow-measurement regime when $\Gamma\ll|\Omega_S|$, $S_{PP,\t{ext}}$ comes from the thermal noise of fast-decaying modes (see \secref{sec:si:fit}), and in the fast-measurement regime when $\Gamma\sim|\Omega_S|$, $S_{PP,\t{ext}}=0$. There is no detectable extraneous noise in the $X$ quadrature of light in our experiments.
The spectrum of the imprecision noise for measurements on the $P$ quadrature is given by
\begin{equation}
S_\t{imp}[\Omega]=\frac{1/4+S_{PP,\t{ext}}[\Omega]}{\eta\Gamma},
\end{equation}
The spectrum of the backaction noise is given by
$S_\t{BA}[\Omega]=\hbar^2\left(\Gamma(1+\zeta^2)+\gamma_\t{sc}\right)$,
where $\gamma_\t{sc}$ is the decoherence rate of the oscillator due to spontaneous scattering, which is proportional to the probe power. We conservatively estimate $\gamma_\t{sc}/\Gamma$ as $1/\mathcal{C}_\t{q}$ (as if all the decoherence of spin oscillators comes from spontaneous scattering).
Overall, the backaction-imprecision product in terms of the two-sided spectral densities is found as
\begin{equation}
\sqrt{S_\t{imp}\,S_\t{BA}}=(\hbar/2)\sqrt{\frac{1}{\eta}\left(1+\frac{S_{PP,\t{ext}}}{\t{SN}}\right)\left(1+\zeta^2+\frac{1}{\mathcal{C}_\t{q}}\right)}.
\end{equation}
where $\t{SN}=1/4$ is the shot noise level. This expression exposes how various imperfections of the measurements, including the finite detection efficiency, the extraneous noise, the ``heating'' due to spontaneous scattering, and the dynamical backaction, elevate the backaction-imprecision product above the quantum limit of $\hbar/2$ in our experiments.

\section{The modeling of the experimental data}\label{sec:si:fit}

To process the experimental data, we model the homodyne spectrum as arising from the dynamics of several oscillator modes coupled to the probe field, using the input-output relations that are expressed analogously to Eqs.(\ref{eq:si:XI-O}) and (\ref{eq:si:PI-O}),
\begin{align}
&\hat{X}_\t{L}^\t{out}(t)=\hat{X}_\t{L}^\t{in}(t)-\sum_{i=1}^{n_\t{modes}}\zeta_i\sqrt{\Gamma_i}\hat{P}_{i}(t),
&\hat{P}_\t{L}^\t{out}(t)=\hat{P}_\t{L}^\t{in}(t)+\sum_{i=1}^{n_\t{modes}}\sqrt{\Gamma_i}\hat{X}_{i}(t).\label{eq:si:XPI-Omodel}
\end{align}
and the Heisenberg equations of motion analogous to Eqs.(\ref{eq:si:XHhc}) and (\ref{eq:si:PHhc}),
\begin{align}
&\frac{d}{dt}\hat{X}_i(t)=\Omega_i \hat{P}_i(t)-\frac{\gamma_{0,i}}{2}\hat{X}_i(t)-\sum_{j=1}^{n_\t{modes}}\zeta_i\sqrt{\Gamma_i\Gamma_j}\hat{X}_{j}(t)-2\zeta_i\sqrt{\Gamma_i}\hat{P}_\t{L}^\t{in}(t)+\hat{F}_i^{X}(t),\label{eq:si:XHhcModel}\\
&\frac{d}{dt}\hat{P}_i(t)=-\Omega_i \hat{X}_i(t)-\frac{\gamma_{0,i}}{2}\hat{P}_i(t)-\sum_{j=1}^{n_\t{modes}}\zeta_j\sqrt{\Gamma_i\Gamma_j}\hat{P}_{j}(t)+2\sqrt{\Gamma_i}\hat{X}_\t{L}^\t{in}(t)+\hat{F}_i^{P}(t).\label{eq:si:PHhcModel}
\end{align}
The index $i$ counts the modes of the model, corresponding to the hybridized resonances we observe in the experimental spectra.
The model accounts for the intrinsic dissipation of the modes characterized by the damping rates $\gamma_{0,i}$, and thermal forces $\hat{F}_i^{X,P}(t)$ via the quantum Langevin approach. The correlators of the thermal forces are
\begin{align}
&\left\langle\hat{F}_i^{X}(t_1)\hat{F}_j^{X}(t_2)\right\rangle=\left\langle\hat{F}_i^{P}(t_1)\hat{F}_j^{P}(t_2)\right\rangle=\delta_{ij}\gamma_{0,i}(n_\t{th}+1/2)\delta(t_1-t_2),
&\left\langle\hat{F}_i^{X}(t_1)\hat{F}_j^{P}(t_2)+\hat{F}_j^{P}(t_2)\hat{F}_i^{X}(t_1)\right\rangle=0.
\end{align}
The intrinsic dissipation in our experiments is dominated by spin depolarization due to the atomic collisions and spontaneous scattering of probe photons, which is why we assume that it symmetrically affects $X$ and $P$, and that the thermal noises are delta-correlated \cite{vasilyev_quantum_2012}. The thermal occupancy of the intrinsic bath is $n_\t{th}=0.9\pm 0.1$, as extracted from the equilibrium macroscopic population distribution of atoms over the magnetic sublevels.

The fast-decaying modes are treated as one, because their frequency splitting is much smaller than their decoherence rates. This mode is accounted differently for different detunings of the optical probe. At large detunings, the measurement rate for the fast-decaying mode also is much smaller than its decoherence rate, and the dynamic backaction is negligible. In this case, it contributes incoherent thermal noise to the measurement of slow-decaying modes. The spectrum of this noise in the $\hat{P}_\t{L}$ quadrature of the output light is given by
\begin{equation}
S_{PP,\t{ext}}[\Omega] = \Gamma'\int_{-\infty}^{\infty} e^{i(\Omega-\Omega_S) \tau} \frac{\langle \Delta g(t+\tau) \Delta g(t) \rangle_c}{\langle \Delta g(t)^2\rangle_c}d\tau,
\end{equation}
where $\Gamma'$ is the measurement rate of the mode and $\langle \Delta g(t+\tau) \Delta g(t) \rangle_c$ is the correlation function of the atomic motion (introduced in \secref{sec:si:spinOscillators}).
Experimentally, we find that this spectrum at frequencies close to the resonance has a Gaussian shape (consistent with a non-Markovian thermal bath), and describe it using the expression
\begin{equation}
S_{PP,\t{ext}}[\Omega]/\mathrm{SN} = A_{b}\,e^{-(\Omega-\Omega_\mathrm{S})^2/(2\gamma^2_{b})},
\end{equation}
where $A_{b}$ is the magnitude of the added noise on resonance in shot noise (SN) units, and $\gamma_{b}$ is the characteristic decay rate.
The spectral width of the broadband noise is closely related to the transition time $\tau$ of atoms through the probe beam, $\gamma_{\text{b}}\sim 1/\tau = v_{\t{th}}/w$, where $w$ is the width of the beam, $v_{\t{th}}=\sqrt{2k_\t{B} T/M_\t{Cs}}\approx\SI{200}{m/s}$ is the thermal velocity atoms, $T=\SI{52}{\degreeCelsius}$ is the operating temperature, $k_\t{B}$ is the Boltzmann constant and $M_\t{Cs}$ is the mass of one atom.

At the detuning of the optical probe equal to \SI{0.7}{GHz}, at which the measurement rate of the spin reaches the oscillation frequency, the fast-decaying mode of the atomic ensemble is in the backaction-dominated regime. We therefore include it as an extra oscillator in Eqs.~(\ref{eq:si:XPI-Omodel}---\ref{eq:si:PHhcModel}). This approach effectively approximates the correlation function of the thermal motion of the mode by an exponential, which in the spectral domain may introduce an error in the frequency window of several hundreds of kHz around the resonance, much smaller than the full bandwidth of the fit (several MHz).

\begin{figure*}[t]
\centering
    \includegraphics{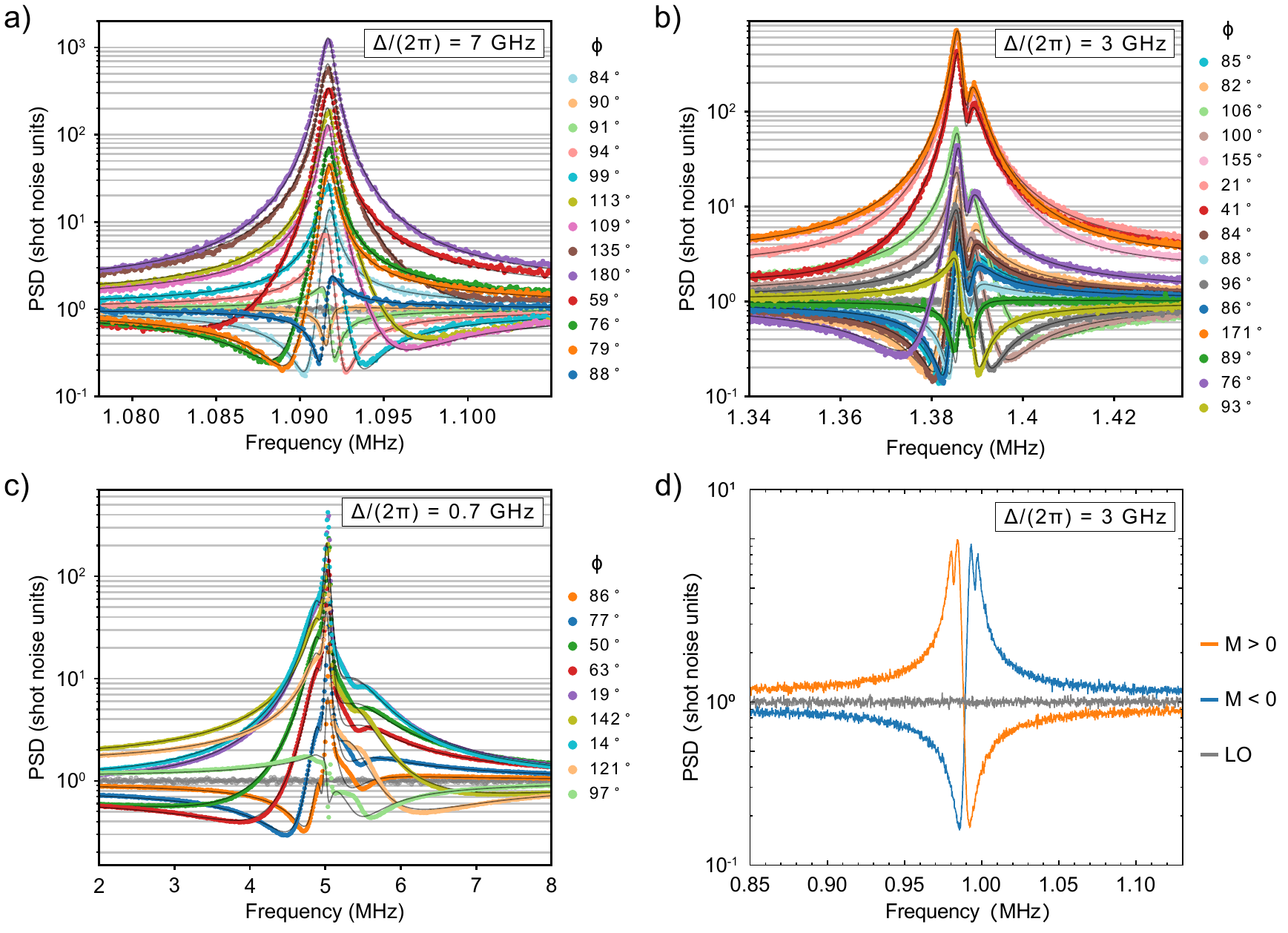}
	\caption{\small{
	a)-c) Power spectral densities (PSD) of homodyne signals recorded at different quadrature angles $\phi$ and laser detunings $\Delta$. The points of different colors show the experimental spectra for different quadrature angles as labeled in the legends. The black curves show the results of global fits at each detuning performed as described in \secref{sec:si:fit}. Gray points show the local oscillator shot noise. Panel a) displays only part of the 17 traces fitted in total.
	d) The effect of changing the oscillator mass, $M$, on the homodyne spectrum measured at a quadrature intermediate between $\hat{X}_\t{L}$ and $\hat{P}_\t{L}$. The blue curve shows the spectrum recorded in a negative mass ($M$) configuration, the orange curve shows the spectrum recorded in a positive mass configuration, and the gray curve shows the local oscillator (LO) shot noise. The sign of the mass was changed by inverting the direction of the magnetic field with respect to the $x$ axis. The spectra were recorded using a \SI{12}{mW} probe detuned from the optical transition by \SI{3}{GHz}.}}
	\label{fig:si:fullFits}
\end{figure*}

The full comparison between the model and the experimental data at different optical detunings is shown in \figref{fig:si:fullFits}. The data obtained at \SI{7}{GHz} optical detuning is described by the response of a single oscillator mode to the measurement backaction. The data obtained at \SI{3}{GHz} detuning is described with $n_\t{modes}=2$. At \SI{0.7}{GHz}, we include the fast-decaying mode in the model and describe the experiment with $n_\t{modes}=3$.
The homodyne spectra at all quadratures are processed in one global fit, where the resonance frequencies $\Omega_i$, the measurement rates $\Gamma_i$, the dynamical backaction coefficients $\zeta_i$, the intrinsic damping rates $\gamma_{0,i}$, and the quadrature angles $\phi$ are free parameters, and the values of the thermal occupancy $n_\t{th}$ and the detection efficiency $\eta$ are taken from independent calibrations.
When processing the broadband measurements at \SI{0.7}{GHz}, we additionally correct for the frequency response of the measurement electronic chain. The total quantum cooperativity for the data in \figref{fig:si:fullFits}c is 4.6. 

\section{The sign of the mass}\label{sec:si:massFlip}

Spin oscillators can have positive or negative effective masses depending on the orientation of the mean spin alignment $\langle\hat{J}_x\rangle$ with respect to the magnetic field. The sign of the mass determines the overall sign of the response $\chi[\Omega]$ of the oscillator to generalized forces, including the quantum backaction force when the oscillator is subjected to linear measurements. Negative-mass oscillators can cancel measurement backaction on regular material oscillators~\cite{moller_quantum_2017}, and become entangled with them~\cite{thomas_entanglement_2020}.

The sign of the oscillator mass, together with the detection angle and the Fourier frequency, determines the sign of the backaction-imprecision correlations observed in homodyne measurement records.
For multiple resonances, it also inverts the signs of the frequency splittings due to the Stark and quadratic Zeeman effects.
The total effect of inverting the mass sign on homodyne spectra is therefore the reflection of the spectra with respect to the Larmor frequency.
We observe this in \figref{fig:si:fullFits}d, where we invert the sign of the mass by changing the direction of the magnetic field.

\section{The spectrum of the homodyne signal in the presence of dynamical backaction}\label{sec:si:optDamping}

To illustrate the effect that the deviation of the interaction Hamiltonian from pure position measurement-type ($\zeta=0$) has on the detected spectra and the squeezing of light, we present an analytical solution for the optimum-quadrature homodyne spectrum in the single-oscillator model with arbitrary $\zeta\in[-1,1]$ under the rotating-wave approximation (RWA).  
For a single mode, by solving Eqs.~(\ref{eq:si:XHhcModel}-\ref{eq:si:PHhcModel}) and using the input-output relations given by \eqref{eq:si:XPI-Omodel}, we find the spectrum of the output signal neglecting the detection losses as
\begin{equation}\label{eq:si:spectrumRWA}
S_\phi[\Omega]/\t{SN}= 1+2\t{Re}\left[\mathcal{A} \chi[\Omega]\right]+|\mathcal{A}\chi[\Omega]|^2\left(1+\frac{\gamma_\t{th}+\gamma_0}{\Gamma(1+\zeta)}\right),
\end{equation}
where $\t{SN}=1/4$ is the shot noise level, $\chi[\omega]=-(1/2)/(\Delta\Omega+i\gamma/2)$ is the RWA force susceptibility, $\Delta\Omega=\Omega-\Omega_\t{S}$ is the Fourier-detuning from the oscillator resonance, $\gamma=\gamma_0+2\zeta\Gamma$ is the total oscillator linewidth, and the transduction factor $\mathcal{A}$ is
\begin{equation}
\mathcal{A}=i \Gamma (1+\zeta) \left((1+\zeta)+(1-\zeta) e^{-2 i \phi }\right).
\end{equation}
By minimizing \eqref{eq:si:spectrumRWA} over the quadrature angle $\phi$, we find the frequency-dependent maximum-squeezing angle $\phi_\t{min}$ via
\begin{equation}
\tan\left(2\phi_\t{min}[\Omega]\right)=-\frac{2\Delta\Omega}{\gamma_\t{dec}},
\end{equation}
where the total decoherence rate $\gamma_\t{dec}=\gamma
_\t{th}+\gamma_\t{QBA}$ is the sum of the decoherence rates due to the intrinsic thermal noise, $\gamma_\t{th}$ and the quantum backaction, $\gamma_\t{QBA}$ which are defined as
\begin{align}
&\gamma_\t{th}=(2n_\t{th}+1)\gamma_0,&
&\gamma_\t{QBA}=\Gamma(1+\zeta^2).
\end{align}
The shot-noise normalized signal spectrum at the optimum quadrature is
\begin{equation}
S_{\phi_\t{min}}[\Omega]/\t{SN}=1-\frac{2\gamma_\t{DBA}/\gamma}{1+(2\Delta\Omega/\gamma)^2}-\frac{2\gamma_\t{dec}\Gamma/\gamma^2}{1+(2\Delta\Omega/\gamma)^2}\left((1-\zeta^2)\sqrt{1+\left(\frac{2\Delta\Omega}{\gamma_\t{dec}}\right)^2}-(1+\zeta^2)\right),
\end{equation}
where $\gamma_\t{DBA}=2\zeta\Gamma$ is the contribution of the dynamical backaction to the total oscillator linewidth (the optical damping).
The absolute minimum of the spectrum is found by further minimizing $S_{\phi_\t{min}}[\Omega]$ over $\Delta\Omega$, which can be done analytically in the general case, but yields a cumbersome result. 
Instead of presenting this result, we restrict the attention to the case $\zeta\ll1$, which is relevant to our experiments, and estimate the minimum noise level by evaluating $S_{\phi_\t{min}}[\Omega]$ at $\Delta\Omega_\t{min,\zeta=0}=1/2\sqrt{\gamma(2\gamma_\t{dec}+\gamma)}$, the optimum Fourier detuning for $\zeta=0$. The result is
\begin{equation}
S_\t{min}\approx 1-\frac{\Gamma}{\gamma_\t{dec}+\gamma_0}-\frac{(\gamma_{0}+\gamma_\t{th})\gamma_\t{DBA}}{(\gamma_{0}+\gamma_\t{dec})^2}.
\end{equation}
When the thermal occupancy of the intrinsic bath is close to zero, and the quantum cooperativity is in the intermediate regime, such that $\gamma_\t{dec}$ has the same order of magnitude as $\gamma_0$, there is an improvement in the minimum noise level from a small positive optical damping.

\section{The generation of the collimated tophat beam}\label{sec:si:tophat}

\begin{figure}[h]

        \includegraphics[width=1\textwidth]{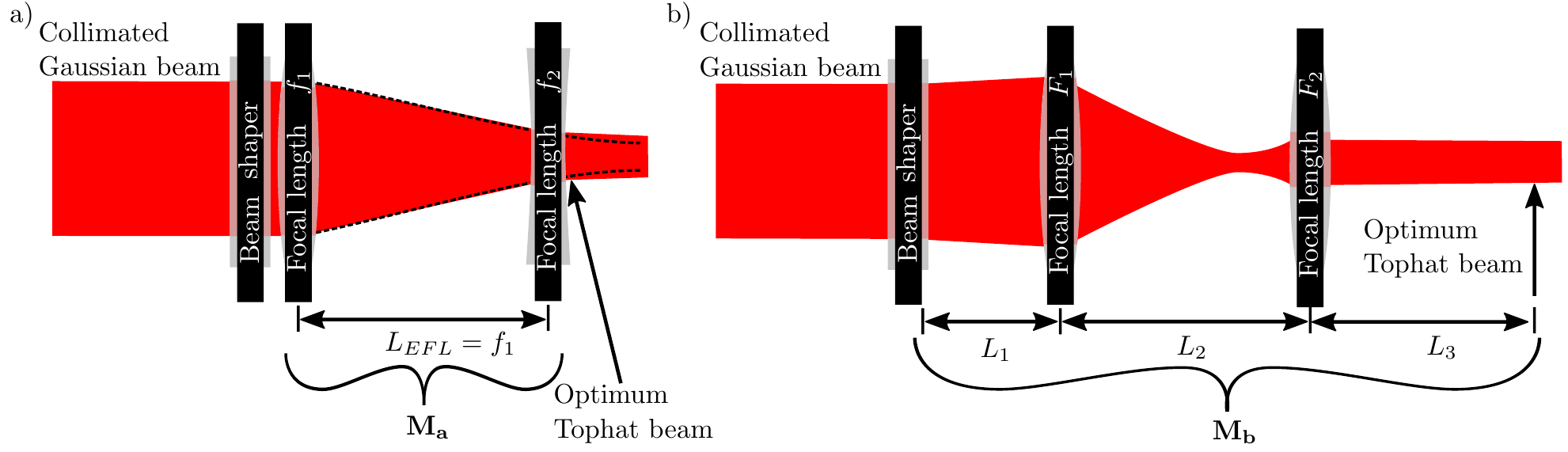}
	\caption{Optical setups for the generation of collimated tophat beams. $\mathbf{M}_{a,b}$ are ray transfer matrices. a) A simple setup. The dashed black line shows how the beam would propagate after passing the beam shaper and the lens $f_1$, but without passing the negative lens $f_2$. EFL: effective focal length. b) A realistic setup designed using the condition $\mathbf{M}_a=\mathbf{M}_b$. Beam shaper: Gaussian-to-tophat beam-shaping lens.}
	\label{fig:si:tophat}
\end{figure}

Optical beams with tophat transverse profiles are commonly produced by passing a collimated Gaussian beam through an aspherical beam shaper, and focusing the beam after the shaper using a spherical lens. In this configuration, the optimum tophat profile (giving the sharpest roll-off of the intensity distribution in the transverse direction) is realized before the focal point, and the beam is tightly focused. In our experiment, it is essential to create a beam in which the tophat profile coincides with the position of the beam waist, and has a relatively large transverse size, enabling a long Rayleigh length extending over the entire cell channel.

An intuition on how to produce a tophat beam that fulfills our criteria can be obtained by examining the setup shown in figure \figref{fig:si:tophat}a, which is a straightforward extension of the usual beam shaper application scheme with an addition of a negative lens $f_2$. 
The optimum tophat transverse profile is realized at a distance one effective focal length (EFL) away from the first lens. The transverse width is proportional to the focal length $f_1$. 
The beam is converging at the optimum point, because of the full fan angle of the tophat beam shaper (i.e. the divergence the shaper introduces in the beam). 
By placing an appropriate negative lens $f_2$ in the optimum point, the beam can be collimated, and its waist position made coincide with the optimum location of the transverse profile. 
The required focal length of the negative lens can be calculated given the size of the input Gaussian beam, $w_\text{in}$, and the full fan angle of the beam shaper, $\phi_\text{FA}$, as $f_2 = \frac{\phi_\text{FA}/w_\text{in} f_1}{\phi_\text{FA}/w_\text{in}-1/f_1}$.

The setup in \figref{fig:si:tophat}a would be challenging to implement directly, because the waist position of the beam is located inside the cell, where placing a lens is hardly realistic. 
However, one can find an optical setup with an identical ray transfer matrix to the one in \figref{fig:si:tophat}a, but realized using a different physical arrangement of lenses. Such a setup is shown in \figref{fig:si:tophat}b.
The transfer matrices for the two setups, $\mathbf{M}_a$ and $\mathbf{M}_b$, are given by
\begin{align}
    \mathbf{M}_a &=
    \mathbf{L}(f_2)\mathbf{S}(f_1)\mathbf{L}(f_1),&
    \mathbf{M}_b &=
    \mathbf{S}(L_3)\mathbf{L}(F_2)\mathbf{S}(L_2)\mathbf{L}(F_1)\mathbf{S}(L_1),
\end{align}
where the matrices for propagation in free space, $\mathbf{S}$, and passing through a lens, $\mathbf{L}$, respectively, are
\begin{align}
    \mathbf{S}(L) &= 
    \begin{bmatrix}
    1 & L \\
    0 & 1
    \end{bmatrix},&
    \mathbf{L}(f) &= 
    \begin{bmatrix}
    1 & 0 \\
    -1/f & 1
    \end{bmatrix}.
\end{align}
In our experiment, the setup in \figref{fig:si:tophat}b is implemented using lenses of pre-determined focal lengths $F_1$ and $F_2$, while the separating distances $L_1$, $L_2$ and $L_3$ are adjusted to meet the condition $\mathbf{M}_a  = \mathbf{M}_b$. 
Additionally, the matrix $\mathbf{M}_a$ is supplemented by an inversion in the transverse plane, which can be interpreted as passing the beam through an extra 4f optical system, which is done in order to have more flexibility in the choice of lenses and more control over the resulting distances.

\bibliography{references}